\documentclass[pre,twocolumn,english,longbibliography]{revtex4-1} 

\usepackage{babel}
\usepackage{amsmath}
\usepackage{amssymb}
\usepackage{graphicx}
\graphicspath{figs}
\usepackage{epstopdf} 

\usepackage[T1]{fontenc}
\usepackage[utf8x]{inputenc} 
\usepackage{esint}
\usepackage{verbatim}
\usepackage[unicode=true]{hyperref}
\usepackage[capitalize,nameinlink]{cleveref}
\creflabelformat{equation}{#2\textup{#1}#3}
\usepackage[table]{xcolor}
\newcommand{\re}[1]{\textcolor{black}{#1}}
\makeatletter

\pdfpageheight\paperheight
\pdfpagewidth\paperwidth

\usepackage{bbold}

\usepackage{xcolor}
\hypersetup{
    colorlinks,
    linkcolor={red!50!black},
    citecolor={blue!50!black},
    urlcolor={blue!80!black}
}

\usepackage{amsmath}
\DeclareMathOperator*{\argmax}{argmax}

\newcommand{\beginsupplement}{
        \setcounter{table}{0}
        \renewcommand{\thetable}{S\arabic{table}}
        \setcounter{figure}{0}
        \renewcommand{\thefigure}{S\arabic{figure}}
     }

\newcommand{\beqn}{\begin{eqnarray}}
\newcommand{\eeqn}{\end{eqnarray}}
\newcommand{\beq}{\begin{equation}}
\newcommand{\eeq}{\end{equation}}

\renewcommand{\>}{\rangle}

\newcommand{\ch}{}
     
\makeatother

\begin{document}

\title{Inferring the immune response from repertoire sequencing}

\author{Maximilian Puelma Touzel}

\affiliation{Laboratoire de physique de l'\'Ecole normale sup\'erieure
  (PSL University), CNRS, Sorbonne  Universit\'e, Universit\'e de
  Paris, 75005 Paris, France}
\affiliation{Mila, Université de Montréal, Montreal, Canada}

\author{Aleksandra M. Walczak}
\thanks{Corresponding authors. These authors contributed equally.}
\affiliation{Laboratoire de physique de l'\'Ecole normale sup\'erieure
  (PSL University), CNRS, Sorbonne  Universit\'e, Universit\'e de
  Paris, 75005 Paris, France}

\author{Thierry Mora}
\thanks{Corresponding authors. These authors contributed equally.}
\affiliation{Laboratoire de physique de l'\'Ecole normale sup\'erieure
  (PSL University), CNRS, Sorbonne  Universit\'e, Universit\'e de
  Paris, 75005 Paris, France}

\vspace{0.5cm}

\begin{abstract}
High-throughput sequencing of B- and T-cell receptors makes it possible to track immune repertoires across time, in different tissues, and in acute and chronic diseases or in healthy individuals. However, quantitative comparison between repertoires is confounded by variability in the read count of each receptor clonotype due to sampling, library preparation, and expression noise. Here, we present a general Bayesian approach to disentangle repertoire variations from these stochastic effects. Using replicate experiments, we first show how to learn the natural variability of read counts by inferring the distributions of clone sizes as well as an explicit noise model relating true frequencies of clones to their read count. We then use that null model as a baseline to infer a model of clonal expansion from two repertoire time points taken before and after an immune challenge. Applying our approach to yellow fever vaccination as a model of acute infection in humans, we identify candidate clones participating in the response.

\end{abstract}

\maketitle

Next generation sequencing allows us to gain access to repertoire-wide data supporting more comprehensive repertoire analysis and more robust vaccine design \cite{Benichou2011}. 
Despite large-scale efforts \cite{Glanville2017}, how repertoire statistics respond to such acute perturbations is unknown. 
Longitudinal repertoire sequencing (RepSeq) makes possible the characterization of repertoire dynamics. 
Despite the large number of samples (clones) in these datasets lending it to model-based inference, there are few existing model-based approaches to this analysis. 
Most current approaches (e.g. \cite{Chu2019}) quantify repertoire response properties using measurement statistics that are limited to what is observed in the sample, rather than what transpires in the individual.
Model-based approaches, in contrast, can in principle capture features of the actual repertoire response to, for instance ongoing, natural stimuli, modeled as a point process of infections, and giving rise to diffusion-like response dynamics. 
Another regime for model-based approaches is the response to a single, strong perturbation, such as a vaccine, giving rise to a stereotyped, transient response dynamics.
In either case, a measurement model is needed since what is observed (molecule counts) is indirect.
We also only observe a small fraction of the total number of clones, so some extrapolation is necessary. 
Finally, both the underlying clonal population dynamics and the transformation applied by the measurement is stochastic, each contributing its own variability, making inferences based on sample ratios of molecule counts inaccurate.

Inference of frequency variation from sequencing data has been intensely researched in other areas of systems biology, such as in RNAseq studies. There, approaches are becoming standardized (DESEQ2 \cite{Love2014}, EdgeR \cite{Robinson2008}, $etc.$) and technical problems have been formulated and partly addressed.
The differences between RNAseq and RepSeq data, however, means that direct translation of these methods is questionable. Moreover, the known structure of clonal populations may be leveraged for model-based inference using RepSeq, potentially providing advantages over existing RNAseq-based approaches.

Here, we take a generative modeling approach to repertoire dynamics. Our model incorporates known features of clonal frequency statistics and the statistics of the sequencing process. The models we consider are designed to be learnable using RepSeq data, and then used to infer properties of the repertoires of the individuals providing the samples. To guide its development, we have analyzed a longitudinal dataset around yellow fever vaccination (some results of this analysis are published \cite{Pogorelyy12704}). Yellow fever serves as model of acute infection in humans and here we present analyses of this data set that highlights the inferential power of our approach to uncover perturbed repertoire dynamics.

\section*{Results}

\subsection*{Modeling repertoire variation}

\begin{figure*}
\includegraphics[width=\linewidth]{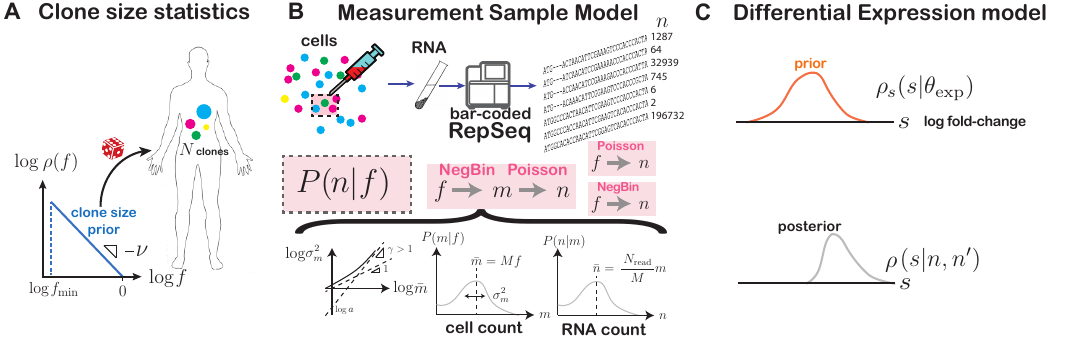}
\centering{}
\caption{
\emph{Model components}. (A) Clone frequencies are sampled from a prior density of power law form with power $\nu$ and minimum frequency, $f_\textrm{min}$. (B) Each clone's frequency $f$ determines the count distribution, $P(n|f)$, that governs its mRNA count statistics in the observed sample. We consider 3 forms for $P(n|f)$: Poisson, negative binomial, and a two-step (negative binomial to Poisson) model. The negative binomial and two-step measurement models are parametrized through a mean-variance relationship specifying the power, $\gamma$, and coefficient, $a$, of the over-dispersion of cell count statistics. The mean cell count scales with the number of cells in the sample, $M$, while the mean read count scales with with the number of cells, $m$, and the sampling efficiency, $M/N_{\textrm{read}}$, with $N_{\textrm{read}}$ the measured number of molecules in the sample. The parameters of the measurement model are learned on pairs of sequenced repertoire replicates.  (C) Differential expression is implemented in the model via a random log fold change, $s$, distributed according to the prior $\rho(s|\theta_\textrm{exp})$. The prior's parameters, $\theta_\textrm{exp}$, are learned from the dataset using maximum likelihood. Once learned, the model is used to compute posteriors over $s$ given observed count pairs, which is used to make inferences about specific clones.
\label{fig:fullmodel}}
\end{figure*}

To describe the stochastic dynamics of an individual clone, we define a probabilistic rule relating its frequency $f'$ at time $t'$ to its frequency $f$ at an earlier time $t$: $G(f',t'|f,t)$. In this paper, $t$ and $t'$ will be pre- and post-vaccination time points, but more general cases may be considered.
It is also useful to define the probability distribution for the clone frequency at time $t$, $\rho(f)$ (Fig.~\ref{fig:fullmodel}A).

The true frequencies of clones are not directly accessible experimentally. Instead, sequencing experiments give us number of reads for each clonotypes, $n$, which is a noisy function of the true frequency $f$, described by the conditional probability $P(n|f)$ (Fig.~\ref{fig:fullmodel}B). Correcting for this noise to uncover the dynamics of clones is essential and is a central focus of this paper.

Our method proceeds in two inference steps, followed by a prediction step. First, using same-day replicates at time $t$, we jointly learn the characteristics of the frequency distribution $\rho(f)$  (Fig.~\ref{fig:fullmodel}A) and the noise model $P(n|f)$ (Fig.~\ref{fig:fullmodel}B). Second, by comparing repertoires between two time points $t$ and $t'$, we infer the parameters of the evolution operator $G(f',t'|f,t)$, using the noise model and frequency distribution learned in the first step (Fig.~\ref{fig:fullmodel}C). Once these two inferences have been performed, the dynamics of individual clones can be estimated by Bayesian posterior inference. These steps are described in the remaining Results sections. In the rest of this section, we define and motivate the classes of model that we chose to parametrize the three building blocks of the model, schematized in Fig.~\ref{fig:fullmodel}: the clone size distribution $\rho(f)$, the noise model $P(n|f)$, and the dynamical model $G(f',t'|f,t)$.

{\ch This method differs from existing approaches of differential expression detection \cite{Love2014,Robinson2008} in at least three ways. First, it can explicitly account for the finite count of cells with a given clonotype. That level of description does not exist in differential expression. Second, it follows a Bayesian approach, which gives the posterior probability of expansion of particular clones, rather than a $p$-value  (although see \cite{Breda2019} for a recent Bayesian approach to differential expression). Third, it includes information about the clone size distribution as a prior to assess the likelihood of expansion, and can thus extract information about clonal structure and diversity. A detailed description of classical differential expression analysis is given in the Methods section.}

\subsubsection*{Distribution of lymphocyte clone sizes}

The distribution of clone sizes in memory or unfractioned TCR repertoires has been observed to follow a power law in human \cite{Mora2016e,Gerritsen_thesis,Greef2019} and mice \cite{Zarnitsyna2013,Heather2017}. These observations justify parametrizing the clone size distribution as
\beq
\rho(f)=Cf^{-\nu}, \qquad f_{\rm min}\leq f<1,
\eeq
and $C$ a normalizing constant. 
We will verify in the next section that this form of clone size distribution describes the data well.
For $\nu>1$, which is the case for actual data, the minimum $f_{\rm min}$ is required to avoid the divergence at $f=0$. This bound also reflects the smallest possible clonal frequencies given by the inverse of the total number of lymphocytes, $1/N_{\rm cell}$. The frequencies of different clones are not independent, as they must sum up to 1: $\sum_{i=1}^Nf_i=1$, where $N$ is the total number of clones in the organism. The joint distribution of frequencies thus reads:
\beq
\rho_N(f_1,\ldots,f_N)\propto\prod_{i=1}^N\rho(f_i)\delta\left(\sum_{i=1}^Nf_i-1\right).
\eeq
This condition, $\sum_{i=1}^Nf_i=1$, will be typically satisfied for large $N$ as long as $\<f\>=\int \textrm{d}f \,f\rho(f) = N^{-1}$ (see Methods), but we will need to enforce it explicitly during the inference procedure. 

\subsubsection*{Noise model for sampling and sequencing}

The noise model captures the variability in the number of sequenced reads as a function of the true frequency of its clonotypes in the considered repertoire or subrepertoire. The simplest and lowest-dispersion noise model assumes random sampling of reads from the distribution of clonotypes. This results in $P(n|f)$ being given by a Poisson distribution of mean $fN_{\rm read}$, where $N_{\rm read}$ is the total number of sequence reads. Note that for the data analyzed in this paper, reads are collapsed by unique barcodes corresponding to individual mRNA molecules. 

Variability in mRNA expression as well as library preparation introduces uncertainty that is far larger than predicted by the Poisson distribution. This motivated us to model the variability in read counts by a negative binomial of mean $\bar n=fN_{\rm read}$ and variance $\bar n+a\bar n^\gamma$, where $a$ and $\gamma$ control the over-dispersion of the noise. Negative binomial distributions were chosen because they allow us to control the mean and variance independently, and reduce to Poisson when $a=0$. These distributions are also popular choices for modeling RNAseq variability in differential expression methods \cite{Robinson2010,Love2014}.

A third noise model was considered to account explicitly for the number of cells representing the clone in the sample, $m$. In this two-step model, $P(m|f)$ is given by a negative binomial distribution of mean $\bar m=fM$ and variance $\bar m+a\bar m^\gamma$, where $M$ is the total number of cells represented in the sample. $P(n|m)$ is a Poisson distribution of mean $mN_{\rm read}/M$. The resulting noise model is then given by $P(n|f)=\sum_m P(n|m)P(m|f)$. The number of sampled cells, $M$, is unknown and is a parameter of the model. Note that this two-step process with the number of cells as an intermediate variable is specific to repertoire sequencing, and has no equivalent in RNAseq differential expression analysis. {\ch The choice of order between the Poisson distribution and the negative binomial is mainly one of tractability.  Ultimately the main motivation for the model is that it performs better empirically (see below).}

\subsubsection*{Dynamical model of the immune response}
Finally, we must specify the dynamical model for the clonal frequencies. In the context of vaccination or infection, it is reasonable to assume that only a fraction $\alpha$ of clones respond \re{by either expanding or contracting}. We also assume that expansion or contraction does not depend on the size of the clone itself. Defining $s=\ln(f'/f)$ as the log-fold factor of expansion or contraction, we define:
\beq
G(f'=fe^s,t'|f,t)\textrm{d}f'=\rho_s(s)\textrm{d}s.
\eeq
with
\beq\label{eq:exp}
\rho_s(s)= (1-\alpha)\delta(s-s_0)+\alpha \rho_{\rm exp}(s-s_0),
\eeq
where $\rho_{\rm exp}$ describes the expansion of responding clones, and $s_0<0$ corresponds to an overall contraction factor ensuring that the normalization of frequencies to 1 is satisfied after expansion. In the following, we shall specialize to particular forms of $\rho_{\rm exp}$ depending on the case at hand.

\subsection*{Inferring the noise profile from replicate experiments} 

To study variations arising from experimental noise, we analysed replicates of repertoire sequencing experiments. The tasks of learning the noise model and the distribution of clone sizes are impossible to dissociate. To infer $P(n|f)$, one needs to get a handle on $f$, which is unobserved, and for which the prior distribution $\rho(f)$ is essential. Conversely, to learn $\rho(f)$ from the read counts $n$, we need to deconvolve the experimental noise, for which $P(n|f)$ is needed. Both can be learned simultaneously from replicate experiments (i.e. $f^\prime=f$), using maximum likelihood estimation. For each clone, the probability of observing $n$ read counts in the first replicate and $n'$ read counts in the second replicate reads:
\beq\label{eq:null}
P(n,n'|\theta_{\rm null})=\int_{f_{\rm min}}^1 \textrm{d}f\, \rho(f|\theta_{\rm null}) P(n|f,\theta_{\rm null})P(n'|f,\theta_{\rm null}),
\eeq
where $\theta_{\rm null}$ is a vector collecting all the parameters of both the noise model and the clone size distribution, namely $\theta_{\rm null}=\{f_{\rm min},\nu\}$ for the Poisson noise model, $\theta_{\rm null}=\{f_{\rm min},\nu,a,\gamma\}$ for the negative binomial noise model, and $\theta_{\rm null}=\{f_{\rm min},\nu,a,\gamma,M\}$ for the two-step noise model.

While Eq.~\ref{eq:null} gives the likelihood of a given read count pair $(n,n')$, we need to correct for the fact that we only observe pairs for which $n+n'>0$. In general,
many clones in the repertoire are small and missed in the acquisition process. In any realization, we expect $n+n'>0$ for only a relatively small number of clones, $N_{\rm obs}\ll N$. Typically, $N_{\rm obs}$ is of order $10^5$, while $N$ is unknown but probably ranges from $10^7$ for mouse to $10^8-10^{10}$ for humans \cite{Qi2014,Lythe2016}. Since we have no experimental access to the unobserved clones ($n=n^{\prime}=0$), we maximize the likelihood of the read count pairs $(n_i,n'_i)$, $i=1,\ldots,N_{\rm obs}$, conditioned on the clones appearing in the sample:
\beq\label{eq:MLE}
\hat\theta_{\rm null}=\argmax_{\theta_{\rm null}} \prod_{i=1}^{N_{\rm obs}} \frac{P(n_i,n'_i|\theta_{\rm null})}{1-P(0,0|\theta_{\rm null})}.
\eeq

While the condition $N\<f\>=1$ ensures normalization on average, we may instead require that normalization be satisfied for the particular realization of the data, by imposing:
\beq
	Z=N	P(0,0)\langle f\rangle_{\rho(f|n+n^{\prime}=0)} + \sum_{i=1}^{N_{\textrm{obs}}}\langle f\rangle_{\rho(f|n_i,n^{\prime}_i)}=1,\label{eq:postnorm}
\eeq
where $N$ is estimated as $N=N_{\textrm{obs}}/(1-P(0,0))$. The first term corresponds to the total frequency of the unseen clones, while the second term corresponds to a sum of the average posterior frequencies of the observed clones. Imposing either Eq.~\ref{eq:postnorm} or $N\<f\>=1$ yielded similar values of the parameter estimates, $\hat\theta_{\rm null}$.

To test the validity of the maximum likelihood estimator, Eq.~\ref{eq:MLE}, we created synthetic data for two replicate sequencing experiments with known parameters $\theta_{\rm null}$ under the two-step noise model, and approximately the same number of reads as in the real data.
To do so efficiently, we developed a sampling protocol that deals with the large number of unobserved clones implicitly (see Methods).
Applying the maximum likelihood estimator to these synthetic data, we correctly inferred the ground truth over a wide range of parameter choices (\cref{fig:SM_reinfer_null}).

\begin{figure*}
\includegraphics[width=\linewidth]{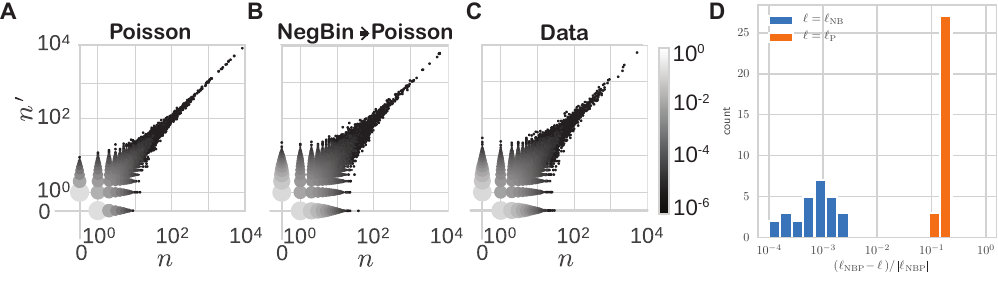}
\centering{}
\caption{
  \emph{Comparison of measurement models}. Pair count distributions sampled from learned (A) negative binomial and (B) Poisson models, compared to (C) data. (D) shows the log likelihoods, $\ell$ (logarithm of the argument of the argmax in Eq.~\ref{eq:MLE}) of the Poisson (P) and negative binomial (NB) models relative to that of the two-step model (NBP). (Example dataset: day-0 replicate pair from donor S2.)  \label{fig:nullstats}
  }
\end{figure*}

\begin{figure}
\includegraphics[width=\linewidth]{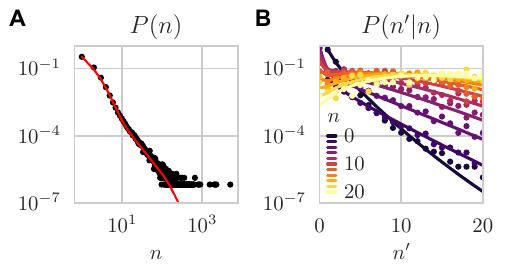}
\centering{}
\caption{
  \emph{Count distributions}. (A) Marginal count distribution, $P(n|\theta_{\textrm{null}})=\sum_{n^{\prime}}P(n,n^{\prime}|\theta_{\textrm{null}})$, and (B) conditional count distribution, $P(n|n^{\prime},\theta_{\textrm{null}})=P(n,n^{\prime}|\theta_{\textrm{null}})/P(n|\theta_{\textrm{null}})$. Both marginal and conditional distributions are quantitatively predicted by the model. Lines are analytic predictions of the learned model. Dots are estimated frequencies. (Same data as \cref{fig:nullstats}; two-step noise model).
\label{fig:modelfit}}
\end{figure}

\begin{figure}[h!]
\includegraphics{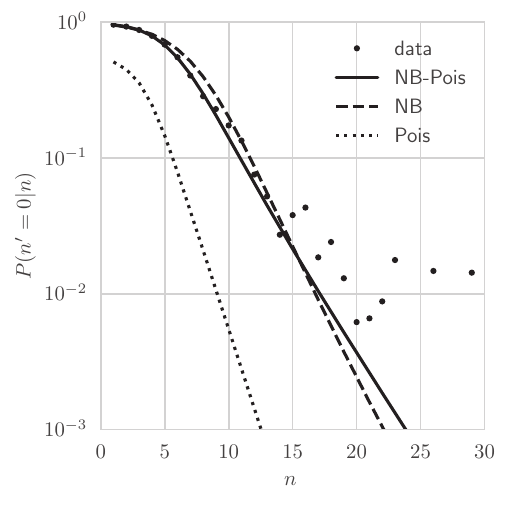}
\centering{}
\caption{
\emph{Dependence of conditional distribution $P(n^\prime=0|n)$ on $n$}. Two-step negative binomial to Poisson model captures tail better than one-step negative binomial model. Poisson model fits poorly. (Example donor S2-day 0 replicate pair.)\label{fig:SM_twostep_better}
}
\end{figure}

Next, we applied the method to replicate sequencing experiments of unfractioned repertoires of 6 donors over 5 time points spanning a 1.5 month period (30 donor-day replicate pairs in total). For a typical pair of replicates, a visual comparison of the $(n,n')$ pairs generated by the Poisson and two-step noise models with the data shows that the Poisson distribution fails to explain the large observed variability between the two replicates, while the two-step model can (Fig.~\ref{fig:nullstats}A-C). The normalized log-likelihood of the two-step model was slightly but significantly higher than that of the negative binomial model, and much larger than that of the Poisson model (\cref{fig:nullstats}D). The two-step model was able to reproduce accurately the distribution of read counts $P(n)$ (Fig.~\ref{fig:modelfit}A), as well as the conditional distribution $P(n'|n)$ (Fig.~\ref{fig:modelfit}B), even though those observables were not explicitly constrained by the fitting procedure. In particular, $P(n)$ inherits the power law of the clone frequency distribution $\rho(f)$, but with deviations at low count numbers due to experimental noise, which agree with the data. 
Also, the two-step model outperformed the negative binomial noise model at describing the long tail of the read count distribution for clones that were not seen in one of the two replicates (see \cref{fig:SM_twostep_better}). 

\begin{figure}
\includegraphics{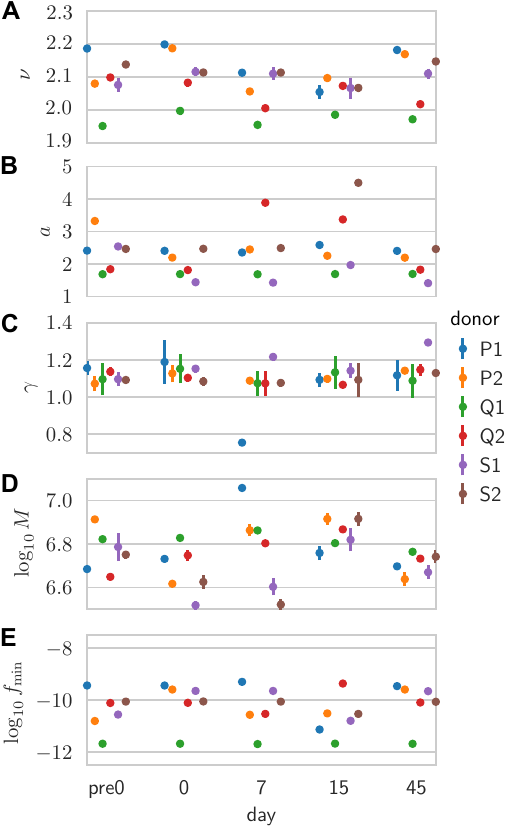}
\centering{}
\caption{
  \emph{Inferred null model parameters}. Inferred values: for (A) the power-law exponent $\nu$ of the clone size distribution; (B) and (C) linear coefficient and exponent of the mean-variance relationship of the noise; (D) effective number of cells; and (E) minimal clonal frequency.
  Each point is inferred from a pair of replicates for a given donor and time point. Error bars are obtained by inverting the Hessian of the log-likelihood projected onto the hyperplane locally satisfying the normalization constraint {\ch (error bars smaller than symbols not visible)}.
\label{fig:nullparas_timeseries}}
\end{figure}

\Cref{fig:nullparas_timeseries} shows the learned values of the parameters for all 30 pairs of replicates across donors and timepoints.
While there is variability across donors and days, {\ch probably due to unknown sources of biological and methodological variability,} there is a surprising degree of consistency. Despite being inferred indirectly from the characteristics of the noise model, estimates for the number of cells in the samples, $M$, are within one order of magnitude of their expected value based on the known concentration of lymphocytes in blood (about one million cells per sample). Likewise, $f_{\rm min}$ is very close to the smallest possible clonal frequency, $1/N_{\rm cell}$, where $N_{\rm cell}\approx 4\cdot 10^{11}$  is the total number of T cells in the organism \cite{Jenkins2010}.

\begin{figure}
\includegraphics[width=\linewidth]{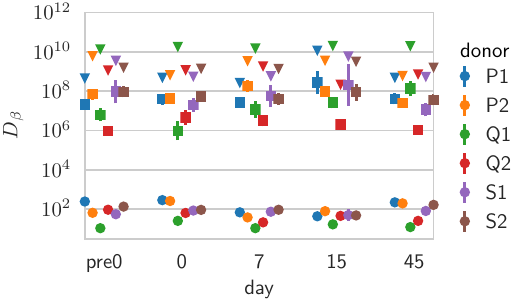}
\centering{}
\caption{\emph{Diversity estimates.} Shown are diversity estimates obtained from the Hill diversities, $D_\beta$, of the inferred clone frequency distributions for $\beta=0$ (estimated total number of clones, $N$), $\beta=1$ (Shannon entropy) and $\beta=2$ (Simpson index), across donors and days. {\ch Error bars reflect parameter uncertainty in the inference, and are computed from the posterior distribution using a Gaussian approximation (error bars smaller than symbols not visible).}
\label{fig:div_estimates}}
\end{figure}

The inferred models can also be used to estimate the diversity of the entire repertoire (observed or unobserved).
The clone frequency distribution, $\rho(f)$, together with the estimate of $N$ can be used to estimate Hill diversities (see Methods):
\beq
D_\beta={\left(\sum_{i=1}^N f_i^\beta\right)}^{\frac{1}{1-\beta}}={\left(N\<f^\beta\>\right)}^{\frac{1}{1-\beta}}.
\eeq
In  \cref{fig:div_estimates}, we show the values, across donor and days, of three different diversities: species richness, i.e. the total number of clones $N$ ($\beta=0$); Shannon diversity, equal to the exponential of the Shannon entropy ($\beta=1$); and Simpson diversity, equal to the inverse probability that two cells belong to the same clone ($\beta=2$). In particular, estimates of $N\approx 10^9$ fall between the lower bound of $10^8$ unique TCRs reported in humans using extrapolation techniques \cite{Qi2014} and theoretical considerations giving upper-bound estimates of $10^{10}$ \cite{Lythe2016} or more \cite{Mora2019}.

\subsection*{Learning the repertoire dynamics from pairs of time points}

Now that the baseline for repertoire variation has been learned from replicates, we can learn something about its dynamics following immunization. The parameters of the expansion model (Eq.~\ref{eq:exp}) can be set based on prior knowledge about the typical fraction of responding clones and effect size. Alternatively, they can be inferred from the data using maximum likelihood estimation (Empirical Bayes approach). We define the likelihood of the read count pairs $(n,n')$ between time points $t$ and $t'$ as:
\beq
\begin{split}\label{eq:fullexp}
  &P_{\rm exp}(n,n'|\theta_{\rm null}, \theta_{\rm exp})=\\
  &\int_{f_{\rm min}}^1 \textrm{d}f\rho(f)\int \textrm{d}s\rho_{s}(s|\theta_{\rm exp})P(n|f, \theta_{\rm null})P(n'|fe^s, \theta_{\rm null}),
  \end{split}
  \eeq
where $\theta_{\rm exp}=\{\alpha,s_0,\bar s\}$ characterizes $\rho_s(s)$ (\cref{eq:exp}) with $\bar s$ parametrizing $\rho_{\textrm{exp}}(s)$, and where $\theta_{\rm null}=\hat \theta_{\rm null}$ is set to the value learned from replicates taken at the first time point $t$.
The maximum likelihood estimator is given by
\beq\label{eq:MLEexp}
\hat\theta_{\rm exp}=\argmax_{\theta_{\rm exp}} \prod_{i=1}^{N_{\rm obs}} \frac{P_{\rm exp}(n_i,n'_i|\hat\theta_{\rm null}, \theta_{\rm exp})}{1-P_{\rm exp}(0,0|\hat\theta_{\rm null}, \theta_{\rm exp})}.
\eeq
This maximization was performed via gradient-based methods. In Methods we give an example of an alternative semi-analytic approach to finding the optimum using the expectation maximization algorithm. 

In addition to normalization at $t$, we also need to impose normalization at $t'$:
\beq
Z'=N	P(0,0)\langle f'\rangle_{\rho(f'|n+n^{\prime}=0)} + \sum_{i=1}^{N_{\textrm{obs}}}\langle f'\rangle_{\rho'(f'|n_i,n^{\prime}_i)},
\eeq
with $\rho(f'|n,n')\propto \int \textrm{d}f\rho(f)G(f'|f)P(n|f)P(n'|f')$ is the posterior distribution of the $f'$ given the read count pair. In practice, we impose $Z=Z'$, where $Z$ is the normalization of the first time point given by Eq.~\ref{eq:postnorm}.
Intuitively, this normalization constraint sets $s_0$ so that the expansion of a few clones is compensated by the slight contraction of all clones.

\begin{figure*}
\includegraphics[width=\linewidth]{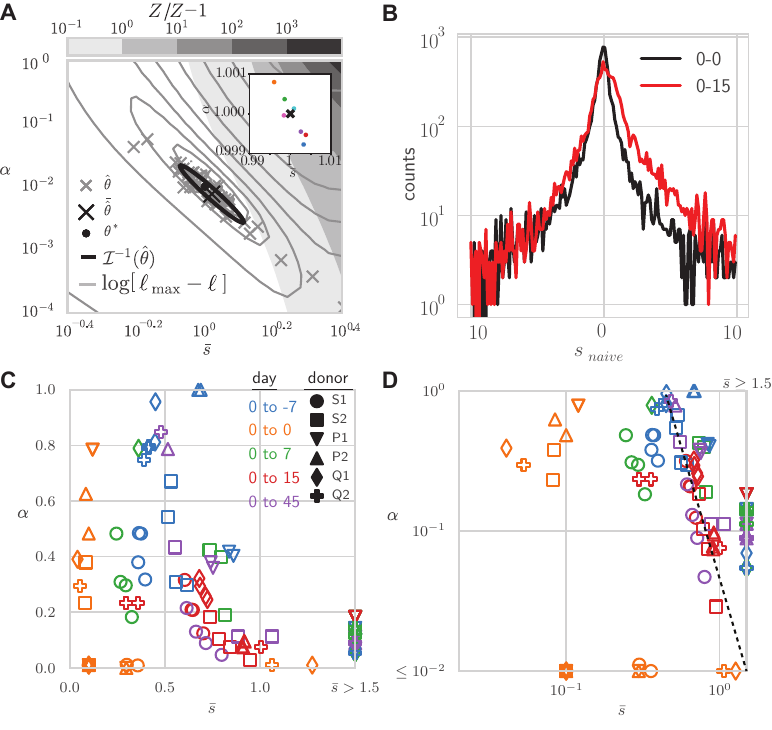}
\centering{}
\caption{
\emph{Inference of clonal expansion on synthetic and real data.}
(A) {\ch Robustness of the} re-inference of the expansion parameters from synthetic data generated with value $\theta_{\rm exp}^*=(\bar{s}^*,\alpha^* )=(1.0,10^{-2})$ (black dot). {\ch Robustness is illustrated in three different ways: 1)
scatter of the re-inferred $\hat\theta_{\rm exp}$ (obtained by maximum likelihood) for 50 realizations (gray crosses, average shown by black cross); 2) isocontour lines for the log-likelihood from one realization (gray contours lines); and 3) ellipse representing the expected variance from one realization, obtained from the inverse Fisher information, $\mathcal{I}^{-1}$ (black line). In addition, gray scale contour regions increasing to the upper-right denote $Z^\prime/Z-1$, the excess in the used normalization} ($\nu=2$,  $f_{\textrm{min}}$ satisfying $N\langle f\rangle_{\rho(f)}$=1; for mouse-sized repertoire parameters: $N=10^6$, $N_{\textrm{reads}}=10^{4}$. Inset shows result for human-sized repertoire ($N=10^9$, $N_{\textrm{reads}}=10^{6}$). 
(B) Empirical histograms of naive log-frequency fold-change $s_{\rm naive}=\ln(n'/n)$. For example data: day-0/day-0 and day-0/day-15 pair comparisons averaged over donors. 
(C) Application to yellow fever vaccination data. Optimal values of $\alpha$ and $\bar{s}$ across all 6 donors and days relative to the day of vaccination (day 0).
{\ch Each pair of different time points allows for 4 comparisons thanks to replicates. Same-day comparisons allow for 2 comparisons depending on which replicate is used as reference.}
(D) Same data from (C) plotted on logarithmic scales for reference. Comparisons with days other than 0 fall on straight line ({\ch guide to the eye}, dashed line).
\label{fig:diffexpr_ex1}
}
\end{figure*}

We first tested the method on synthetic data generated with the expansion model of Eq.~\ref{eq:fullexp}, with an exponentially distributed effect size for the expansion with scale parameter, $\bar{s}$:
\beq\label{eq:onesidedexp}
\rho_{\rm exp}(s')=\frac{1}{\bar s}e^{-s'/\bar s}\Theta(s'),
\eeq
where $\Theta(s')=1$ if $s'>0$ and $0$ otherwise.
We simulated small, mouse-like and large, human-like repertoires (number of clones, $N=10^6$ and $N=10^9$; number of reads/sample $N_{\textrm{reads}}=10^4$ and $N_{\textrm{reads}}=2\cdot 10^6$, respectively), using $\nu=2$ and $f_{\textrm{min}}$ satisfying $N\langle f\rangle_{\rho(f)}$=1. {\ch The procedure consisted of sampling frequencies and log fold factors $N$ times, normalizing by the empirical sum, and then sampling reads from the corresponding measurement distributions, $P(n|f)$ (see Methods for details). Inference on these data produced a pair of estimates $(\bar{s}^*,\alpha^*)$.} For the parameter-free Poisson measurement model, we analyzed the differential expression model, \cref{eq:fullexp}, over a range of biologically plausible parameter values.  
In \cref{fig:diffexpr_ex1}A, we show the parameter space of the inference {from two time points of a \ch single} mouse repertoire generated with $(\bar{s}^*,\alpha^* )=(1.0,10^{-2})$ and $s_0=s_0(\alpha,\bar s)$ fixed by the normalization constraint $Z^\prime=Z$. The sampling procedure was repeated and the set of inferred pair estimates were plotted. The errors are distributed according to a diagonally elongated ellipse (or `ridge'), with a covariance following the inverse of the Hessian of the log-likelihood.
The imprecision of the parameter estimates is due to the small number of sampled responding clones. With
$\alpha^*=0.01$ and $N_{\rm obs}\approx 10^4$ sampled clones, only a few dozens responding clones are detected. For human-sized repertoires, millions of clones are sampled, which makes the inference much more precise (see \cref{fig:diffexpr_ex1}A, inset).

Once learned, the model can be used to compute the posterior probability of a given expansion factor by marginalizing $f$, and using Bayes' rule, 
\begin{align}\label{eq:post}
	\rho(s|n,n^{\prime})\propto \rho_s(s)\int P(n|f)P(n^{\prime}|fe^s)\rho(f)\textrm{d}f.
\end{align}
We illustrate different posterior shapes from synthetic data as a function of the observed count pairs in \cref{fig:posteriors}. We see for instance that the width of the posterior narrows when counts are both large, and that  the model ascribes a fold-change of $s_0$ to clones with $n^{\prime} \lessapprox n$.

Note that the value of the true responding fraction $\alpha$ is correctly learned from our procedure, regardless of our ability to tell with perfect certainty which particular clones responded. By contrast, a direct estimate of the responding fraction from the number of significantly responding clones, as determined by differential expression software such as EdgeR \cite{Robinson2010}, is likely to misestimate that fraction. We applied EdgeR {\ch (see Methods)} to a synthetic repertoire of $N=10^9$ clones, a fraction $\alpha=0.01$ of which responded with mean effect $\bar s=1$, and sampled with $N_{\rm read}=10^6$. EdgeR found $6,880$ significantly responding clones (corrected p-value 0.05) out of $N_{\rm obs}=1,995,139 $, i.e. a responding fraction $6,880/1,995,139 \approx 3\cdot 10^{-3}$ of the observed repertoire, and a responding fraction $6,880/10^9\approx 7\cdot 10^{-6}$ of the total repertoire, underestimating the true fraction $\alpha=10^{-2}$.

\subsection*{Inference of the immune response following immunization} \label{sec:diffexpr}

Next, we ran the inference procedure on sequences obtained from human blood samples across time points following yellow fever vaccination. To guide the choice of prior for $s$, we plotted the histograms of the naive log fold-change $\ln n^{\prime}/n$ (\cref{fig:diffexpr_ex1}B). These distributions show symmetric exponential tails, {\ch although we should recall that these are likely dominated by measurement and sampling noise}. Yet, the {\ch difference between the pair of replicates (black) and the pre- and post-vaccination timepoints (red)} motivates us to model the statistics of expansion factors as:
\beq\label{eq:symmexp}
\rho_{\rm exp}(s)=\frac{1}{2\bar s}e^{-|s|/\bar s},
\eeq
with typical effect size $\bar{s}$. {\ch We also tested other forms of the prior (asymmetric exponential, centered and off-centered Gaussian), but they all yielded lower likelihoods of the data (\cref{tab:prior})}.

We applied the inference procedure (Eq.~\ref{eq:MLEexp}) between the repertoires taken the day of vaccination (day 0), and at one of the other time points (day -7, day 7, day 15, and day 45) after vaccination. Since there are two replicates at each time point, we can make 4 comparisons between any pair of time points. The results are shown in Fig.~\ref{fig:diffexpr_ex1}C {\ch and Fig.~\ref{fig:diffexpr_ex1}C in log-scale}.

Same-day comparisons (day 0 vs day 0) gave effectively zero mean effect sizes ($\bar s<0.1$, below the discretization step of the integration procedure), {\ch or equivalently $\alpha\approx 0$}, as expected. Comparisons with other days yielded inferred values of $\alpha$ and $\bar s$ {\ch mostly} distributed along the same  `ridge', as observed on synthetic data (\cref{fig:diffexpr_ex1}A), {\ch with variations across replicates and donors}. The mean effect size $\bar s$ is highest at day 15, where the peak of the response occurs, but is also substantially different from 0 at all time points except day 0 (including before vaccination at day $-7$), with often high values of $\alpha$. We speculate that these fluctuations reflect natural variations of the repertoire across time, experimental batch effects, {\ch as well as biological variability due to differences in the affinities and precursor frequencies of responding clonotypes}. As a consequence of the natural diversity, 
values of the responding fraction $\alpha$ are not learned with great precision, as can be seen from the variability across the 4 choices of replicate pair, and are probably gross overestimations of the true probability that a naive T cell responds to an infection, which is believed to be of order $10^{-5}-10^{-3}$ \cite{Boer1993}.

\subsection*{Identifying responding clones}

The posterior probability on expansion factors $\rho(s|n,n')$ (Eq.~\ref{eq:post}) can be used to
study the fate and dynamics of particular clones. For instance, we can
identify responding clones as having a low posterior probability of being not expanding $P_{\rm null}=P(s\leq 0|n,n')<0.025$. $P_{\rm null}$ is the Bayesian counterpart of a p-value but differs from it in a fundamental way: it gives the probability that expansion happened given the observations, when a p-value would give the probability of the observations in absence of expansion.  We can define a similar criterion for contracting clones.

\begin{figure*}
\includegraphics[width=\linewidth]{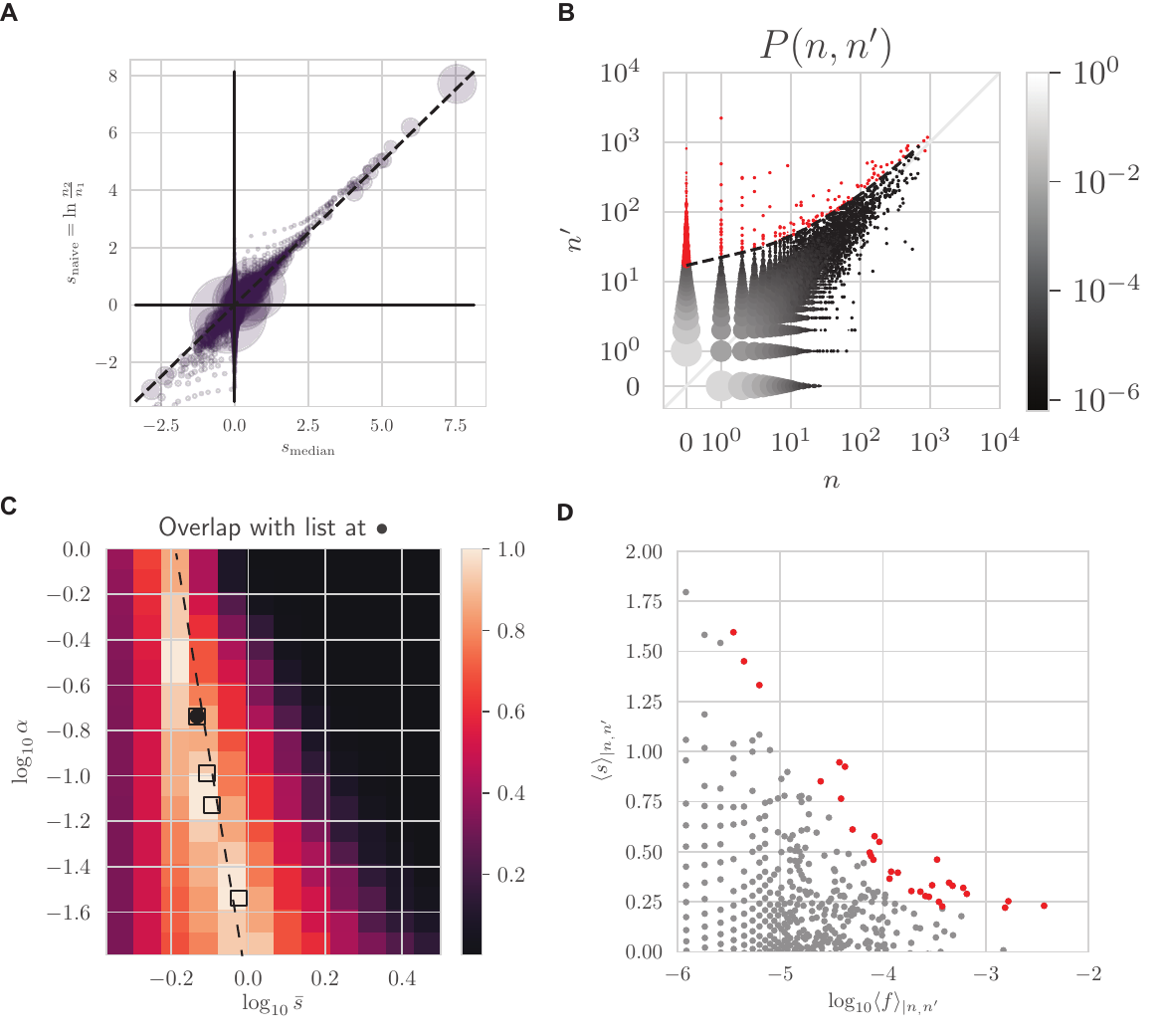}
\centering{}
\caption{
\emph{Identifying responding clones}. (A) Summary statistics of log-frequency fold-change posterior distributions. Comparison of the posterior median log-frequency fold-change and the naive estimate, $\log n^{\prime}/n$ (across clones with $n,n^{\prime}>0$). Each circle is a $(n,n^\prime)$ pair with size proportional to pair count average $(n+n^\prime)/2$. (B) The same threshold for significant expansion in $(n,n^\prime)$-space with identified clones highlighted in red. (C) The optimal values of $\alpha$ and $\bar{s}$ for donor S2 and day-0 day-15 comparison for 3 replicates (square markers). The background heat map is the list overlap (the size of the intersection of the two lists divided by the size of their union) between a reference list obtained at the optimal $\hat\theta_{\rm exp}$ (black dot) and lists obtained at non-optimal $\theta_{\rm exp}$. (D) Mean posterior log fold-change $\<s\>_{\rho(s|n,n')}$ as a function of precursor frequency.
\label{fig:volcano}}
\end{figure*}

To get the expansion or contraction factor of each clone, we can compute the posterior average and median, $\<s\>_{n,n'}=\int \textrm{d}s\,s \rho(s|n,n')$ and $s_{\rm median}$ ($F(s_{\rm median}|n,n')=0.5$, for the cumulative density function, $F(s|n,n')=\int_{-\infty}^s\rho(\tilde{s}|n,n')\textrm{d}\tilde{s}$),  corresponding to our best estimate for the log fold-change.
In \cref{fig:volcano}A, we show how the median Bayesian estimator differs from the naive estimator $s_{\textrm{naive}}=\ln n^{\prime}/n$. While the two agree for large clones for which relative noise is smaller, the naive estimator over-estimates the magnitude of log fold-changes for small clones because of the noise. The Bayesian estimator accounts for that noise and gives a more conservative and more realistic estimate.

\begin{figure}[h!]
\includegraphics{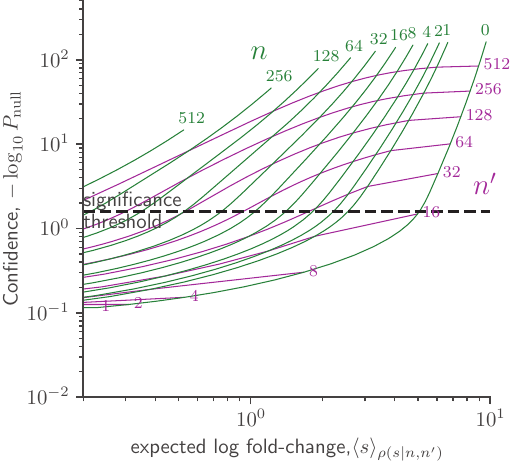}
\caption{
Plot of confidence of expanded response versus average effect size. A significance threshold is placed according to $P_{\textrm{null}}=0.025$, where $P_{\textrm{null}}=P(s\leq 0)$. 
\label{fig:not_valcano}}
\end{figure}

{\ch \Cref{fig:volcano}B shows all count pairs $(n,n')$ between day $0$ and day $15$ following yellow fever vaccination, with red clones above the significance threshold line $P_{\rm null}=0.025$ being identified as responding. Expanded clones can also be read off a plot showing how both $P_{\rm null}$ and $\<s\>_{n,n'}$ vary as one scans values of the count pairs $(n,n')$ (\cref{fig:not_valcano}).}

Given the uncertainty in the expansion model parameters $\theta_{\rm exp}=(\bar s,\alpha)$, we wondered how robust our list of responding clonotypes was to those variations. In \cref{fig:volcano}C, we show the overlap of lists of strictly expanding clones ($P(s\leq 0|n,n')<0.025$) as a function of $\theta_{\rm exp}$, relative to the optimal value $\hat\theta_{\rm exp}$ (black circle). The ridge of high overlap values exactly mirrors the ridge of high likelihood values onto which the learned parameters fall (\cref{fig:diffexpr_ex1}D). Values of $\hat \theta_{\rm exp}$ obtained for other replicate pairs (square symbols) fall onto the same ridge, meaning that these parameters lead to virtually identical lists of candidates for response.

The list of identified responding clones can be used to test hypotheses about the structure of the response. For example, recent work has highlighted a power law relationship between the initial clone size and clones subsequent fold change response in a particular experimental setting \cite{Mayer2019}. We can plot the relationship in our data as the posterior mean log fold change versus the posterior initial frequency, $f$ (\cref{fig:volcano}D). While the relationship is very noisy, emphasizing the diversity of the response, it is consistent with a decreasing dependency of the fold change with the clone size prior to the immune response.

\begin{figure*}[h!]
\includegraphics[width=\linewidth]{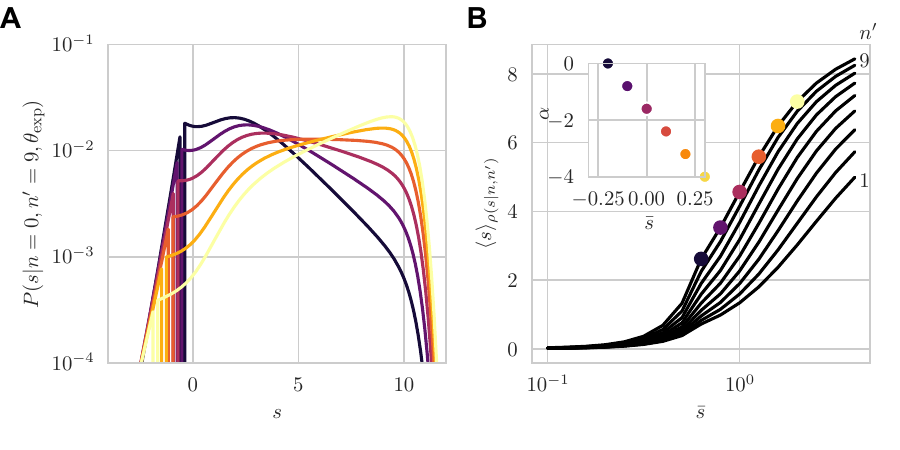}

\caption{
\emph{Competition between $\nu$ and $\bar{s}$ in shaping the posteriors, $\rho(s|0,n^\prime)$}. A) Posteriors for $n^\prime=9$ over a range of $(\bar{s},\alpha)$ pairs spanning the ridge shown in the inset in (B) and \cref{fig:volcano} along which the growth of $\bar{s}$ leads to $\rho(f)$ overwhelming $\rho_s(s)$ as the dominant explanation for observed expansion. (B) The posterior mean versus $\bar{s}$ for values of $n^\prime=1,\dots,9$, with the 5 values of $\bar{s}$ used in (A) shown for $n^\prime=9$.
\label{fig:posteriors}}
\end{figure*}

The robustness of our candidate lists rests on their insensitivity to the details of how the model explains typical expansion. In \cref{fig:posteriors}, we show how the posterior belief varies significantly for count pairs $(0,n^\prime)$, $n^\prime>0$, across a range of values of $\bar{s}$ and $\alpha$ passing along the ridge of plausible models (\cref{fig:volcano}C). A transition from a low to high value of the most probable estimate for $s$ characterizes their shapes and arises as $\bar{s}$ becomes large enough that expansion from frequencies near $f_\textrm{min}$ is plausible, and the dominant mass of clones there makes this the dominate posterior belief. Thus, these posteriors are shaped by $\rho_s(s)$ at low $\bar{s}$, and $\rho(f)$ at high $\bar{s}$. Our lists vary negligibly over this transition, and thus are robust to it.

\section*{Discussion}

Our probabilistic framework describes two sources of variability in clonotype abundance from repertoire sequencing experiments: biological repertoire variations and spurious variations due to noise. We found that in a typical experiment, noise is over-dispersed relative to Poisson sampling noise. This makes the use of classical difference tests such as Fisher's exact test or a chi-squared test inappropriate in this context, and justifies the development of specific methods. {\ch Even in very precise single-cell experiments that do not suffer from expression noise and PCR biases (but are often limited to smaller repertoires owing to high costs), the discrete nature of cell counts creates an irreducible source of Poisson noise. In that case our method would offer a Bayesian alternative to existing approaches.}

As a byproduct, our method learned the properties of the clone size distribution, which is consistent with a power law of exponent $\approx -2.1$ robust across individuals and timepoints, consistent with previous reports \cite{Mora2016e,Gerritsen_thesis,Greef2019}. Using these parameters, various diversity measures could be computed, such as the species richness ($10^8$--$10^9$), which agrees with previous bounds \cite{Qi2014,Lythe2016}, or the ``true diversity'' (the exponential of the Shannon entropy), found to range between $10^6$ and $10^8$. {\ch The inferred null models were found to be conserved across donors and time, indicating that they should be valid for other datasets obtained with the same protocol. This implies that our method could be applicable to situations where replicate experiments are not available, as is often the case. On the other hand, the procedure for learning the null model should be repeated for each distinct protocol using different technologies, using replicate experiments. We applied our method to data from mRNA sequencing experiments, which has the advantage over current DNA immune repertoire sequencing methods of being able to incoorporate unique molecular barcodes. Genomic DNA-based  sequencing does not suffer from expression noise, however the technology is prone to PCR and statistical noise and primer biases. Given that our ultimate choice of noise distributions is often empirically motivated, different modeling choices may be applicable to gDNA datasets.}

The proposed probabilistic model of clonal expansion is described by two parameters: the fraction of clones that respond to the immune challenge, and the typical effect size (log fold-change). While these two parameters were difficult to infer precisely individually, a combination of them could be robustly learned. Despite this ambiguity in the model inference, the list of candidate responding clonotypes is largely insensitive to the parameter details. For clonotypes that rose from very small read counts to large ones, the inferred fold-change expansion factor depended strongly on the priors, and resulted from a delicate balance between the tail of small clones in the clone size distribution and the tail of large expansion events in the distribution of fold-changes.

While similar approaches have been proposed for differential expression analysis of RNA sequencing data \cite{Robinson2008,Robinson2010,Anders2010,Love2014}, the presented  framework was specifically built to address the specific challenges of repertoire sequencing data. Here, the aim is to count proliferating cells, as opposed to evaluating average expression of genes in a population of cells. We specifically describe two steps that translate cell numbers into the observed TCR read counts: random sampling of cells that themselves carry a random number of mRNA molecules, which are also amplified and sampled stochastically. Another difference with previous methods is the explicit Bayesian treatment, which allows us to calculate a posterior probability of expansion, rather than a less interpretable $p$-value.

Here we applied the presented methodology to an acute infection. We have previously shown that it can successfully identify both expanding (from day 0 to 15 after vaccination) and contracting (from day 15 to day 45) clonotypes after administering a yellow fever vaccine. However the procedure is more general and can also be extended to be used in other contexts. For instance, this type of approach could be used to identify response in B-cells during acute infections, by tracking variations in the size of immunoglobulin sequence lineages (instead of clonotypes), using lineage reconstruction methods such as Partis~\cite{Ralph2016a}.
The framework could also be adapted to describe not just expansion, but also switching between different cellular phenoypes during the immune response, {\em e.g.} between the naive, memory, effector memory, {\em etc.} phenotypes, which can obtained by flow-sorting cells before sequencing \cite{Minervina2019}. Another possible application would be to track the clones across different tissues and organs, and detect migrations and local expansions \cite{Kadoki2017}.
The approach requires replicates to quantify natural variability, but this need only be quantified once for the same experimental conditions.  

The proposed framework is not limited to identifying a response during an acute infection, but can also be used as method for learning the dynamics from time dependent data even in the absence of an external stimulus~\cite{Chu2019}. Here we specifically assumed expansion dynamics with strong selection. However, the propagator function can be replaced by a non-biased random walk term, such as genetic drift. In this context the goal is not to identify responding clonotypes but it can be used to discriminate different dynamical models in a way that accounts for different sources of noise inherently present in the experiment. Alternatively, the framework can also be adapted to describe chronic infections such as HIV \cite{Nourmohammad2019}, where expansion events may be less dramatic and more continuous or sparse, as the immune system tries to control the infection over long periods of time.

\section*{Methods}

\subsection*{Code}
All code used to produce the results in this work was custom written in Python 3 and is publicly available online at \url{https://github.com/mptouzel/bayes_diffexpr}.

\subsection*{Normalization of the clonal frequencies}\label{sec:normal}
Here we derive the condition for which the normalization in the joint density is implicitly satisfied. The normalization constant of the joint density is
\begin{equation}
	\mathcal{Z}=\int_{f_\textrm{min}}^1\cdots\int_{f_\textrm{min}}^1\prod_{i=1}^N \rho(f_i)\delta(Z-1)\textrm{d}^N\vec{f} \;\label{eq:joint},
\end{equation}
with $\delta(Z-1)$ being the only factor preventing factorization and explicit normalization. Writing the delta function in its Fourier representation factorizes the single constraint on $\vec{f}$ into $N$ Lagrange multipliers, one for each $f_i$,
\begin{align}
	\delta(Z-1)&=\int_{-i\infty}^{i\infty} \frac{\textrm{d} \mu}{2 \pi}e^{\mu(Z-1)}  \\
	&=\int_{-i\infty}^{i\infty} \frac{\textrm{d} \mu}{2 \pi}e^{-\mu}\prod_{i=1}^N e^{\mu f_i} \;.
\end{align}
Crucially, the multi-clone integral in \cref{eq:joint} over $\vec{f}$ then factorizes. Exchanging the order of the integrations we obtain
\begin{equation}
	\mathcal{Z}=\int_{-i\infty}^{i\infty} \frac{\textrm{d} \mu}{2 \pi} e^{-\mu} \langle e^{\mu f}\rangle^N\;,\label{eq:bigZ}
\end{equation}
with $\langle e^{\mu f}\rangle=\int_{f_\textrm{min}}^1\rho(f)e^{\mu f}\textrm{d}f$. Now define the large deviation function, $I(\mu):=-\frac{\mu}{N}+\log \langle e^{\mu f}\rangle$, so that 
\begin{equation}
	\mathcal{Z}=\int_{-i\infty}^{i\infty} \frac{\textrm{d} \mu}{2 \pi} e^{-N I(\mu)}\;.\label{eq:largedev}
\end{equation}
Note that $I(0)=0$. With $N$ large, this integral is well-approximated by the integrand's value at its saddle point, located at $\mu^*$ satisfying $I'(\mu^*)=0$.  Evaluating the latter gives
\begin{align}
	\frac{1}{N}&=\frac{\langle f e^{\mu^* f}\rangle}{\langle e^{\mu^* f}\rangle}\;.
\end{align} 
If the left-hand side is equal to $\langle f\rangle$, the equality holds only for $\mu^*=0$ since expectations of products of correlated random variables are not generally products of their expectations. 
In this case, we see from \cref{eq:largedev} that $\mathcal{Z}=1$, and so the constraint $N\langle f\rangle=1$ imposes normalization.

\subsection*{Null model sampling}\label{sec:null_sampling}
The procedure for null model sampling is summarized as (1) fix main model parameters, (2) solve for remaining parameters using the normalization constraint, $N \langle f \rangle=1$, and (3) starting with frequencies, sample and use to specify the distribution of the next random variable in the chain.

In detail, we first fix: (a)
the model parameters (e.g. $\{\alpha,a,\gamma,M\}$), excluding $f_{\textrm{min}}$;
(b) the desired size of the full repertoire, $N$;
(c) the sequencing efficiency (average number of UMI per cell), $\epsilon$, for each replicate. 
From the latter we get the mean number of reads per sample, $N^{\textrm{eff}}_{\textrm{reads}}=\epsilon M$.
Note that the actual sampled number of reads is stochastic and so will differ from this fixed value.

We then solve for remaining parameters. Specifically, $f_{\textrm{min}}$ is fixed by the constraint that the average sum of all frequencies, under the assumption that their distribution factorizes, is unity:
\begin{equation}
	N \langle f\rangle_{\rho(f)}=1
\end{equation}
This completes the parameter specification.

We then sample from the corresponding chain of random variables.
Sampling the chain of random variables of the null model can be performed efficiently by only sampling the $N_{\textrm{obs}}=N(1-P(0,0))$ observed clones. This is done separately for each replicate, once conditioned on whether or not the other count is zero. 
Samples with 0 molecule counts can in principle be produced with any number of cells, so cell counts must be marginalized when implementing this constraint. We thus used the conditional probability distributions $P(n|f)=\sum_{m}P(n|m)P(m|f)$ with $m,n=0,1,\dots$. $P(n^\prime|f)$ is defined similarly. Note that these two conditional distributions differ only in their sampling efficiency, $\epsilon$. Together with $\rho(f)$, these distributions form the full joint distribution, which is conditioned on the clone appearing in the sample, i.e. $n+n^{\prime}>0$ (denoted $\mathcal{O}$), 
\begin{align}
	P(n,n^{\prime},f|\mathcal{O})= \frac{P(n|f)P(n^{\prime}|f)\rho(f)}{1-\int{\textrm{d}f \rho(f)\textrm{d}f P(n=0|f)P(n^{\prime}=0|f)}}\;,  
\end{align}
with the renormalization accounting for the fact that $(n,n^{\prime})=(0,0)$ is excluded. The 3 quadrants having a finite count for at least one replicate are denoted $q_{x0}$, $q_{0x}$, and $q_{xx}$, respectively. Their respective weights are
\begin{align}
	P(q_{x0}|\mathcal{O})&=&\sum_{n>0}\int{\textrm{d}f P(n,n^{\prime}=0,f|\mathcal{O})}\;,\\
	P(q_{0x}|\mathcal{O})&=&\sum_{n^{\prime}>0}\int{\textrm{d}f P(n=0,n^{\prime},f|\mathcal{O})}\;,\\
	P(q_{xx}|\mathcal{O})&=&\sum_{\substack{n>0,\\n^{\prime}>0}}\int{\textrm{d}f P(n,n^{\prime},f|\mathcal{O})}.
\end{align}
Conditioning on $\mathcal{O}$ ensures normalization, $P(q_{x0}|\mathcal{O})+P(q_{0x}|\mathcal{O})+P(q_{xx}|\mathcal{O})=1$. Each sampled clone falls in one the three regions according to these probabilities. Their clone frequencies are then drawn conditioned on the respective region, 
\begin{align}
	P(f|q_{x0})&=&\sum_{n>0}P(n,n^{\prime}=0,f|\mathcal{O})/P(q_{x0}|\mathcal{O})\;,\\
	P(f|q_{0x})&=&\sum_{n^{\prime}>0}P(n=0,n^{\prime},f|\mathcal{O})/P(q_{0x}|\mathcal{O})\;,\\
	P(f|q_{xx})&=&\sum_{{n>0,n^{\prime}>0}}P(n,n^{\prime},f|\mathcal{O})/P(q_{xx}|\mathcal{O}).
\end{align}

Using the sampled frequency, a pair of molecule counts for the three quadrants are then sampled as $(n,0)$, $(0,n^{\prime})$, and $(n,n^{\prime})$, respectively, with $n$ and $n^{\prime}$ drawn from the renormalized, finite-count domain of the conditional distributions, $P(n|f,n>0)$. 

\begin{figure}[h!]
\includegraphics{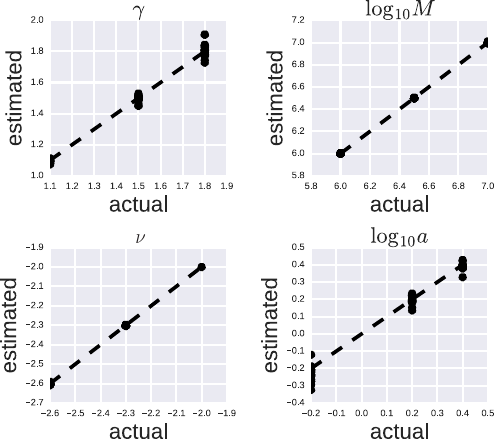}
\centering{}
\caption{
\emph{Reinferring null model parameters}. Shown are the actual and estimated values of the null model parameters used to validate the null model inference procedure over the range exhibited by the data. A 3x3x3x3 grid of points were sampled and results collapsed over each parameter axis. $f_{\rm min}$ was fixed to satisfy the normalization constraint.
}
\label{fig:SM_reinfer_null}
\end{figure}

Using this sampling procedure we demonstrate the validity of the null model and its inference by sampling across the observed range of parameters and re-inferring their values (see \cref{fig:SM_reinfer_null}).

{\ch
\subsection*{Computing Fisher information for constrained maximum likelihood problem}
The replicate model parameters are $\theta=(\nu,a,\gamma,\log_{10}M,\log_{10}f_{\textrm{min}})$. Let $C(\theta)=Z(\theta)-1$ be the constraint equation such that we wish to satisfy $C(\theta)=0$. Let $\theta^*$ denote the parameters maximizing the likelihood subject to $C(\theta)=0$. Then the hyperplane orthogonal to the gradient $\nabla_\theta C (\theta^*)$ and passing through $\theta^*$ is the local subspace in which the constraint is satisfied. The projection of Hessian of the log likelihood, $H$, into this subspace is given by,
\begin{equation}
	\hat{H}=H-PH-HP+PHP
\end{equation}
where the matrix $P=\vec{n}\vec{n}^\top$ projects onto $\vec{n}$, the unit vector co-linear with $\nabla_\theta C (\theta^*)$. The inverse of $H$ has one zero eigenvalue; the remaining eigenvalues characterize the Fisher information at the constrained optimum. Error bars for \cref{fig:nullstats} are the projections of the corresponding ellipsoid onto the respective parameter axes.

When computing error bars for the diversities, we use the standard deviation of the statistics of a Monte Carlo estimate of the log diversities obtained via parameter value samples from the multivariate Gaussian approximation of the likelihood using the projected Hessian, $\hat{H}$.}

{\ch
\subsection*{Comparison to differential expression analysis}
Differential expression deals with RNA-seq data, which reports the bulk expression of a large number of genes in a population of cells, and aims to detect significant differences in expression across different populations, either at different times, or under different conditions.

Repertoire sequencing (RepSeq) and expression analysis aim at inferring fundamentally different quantities, although both do it through the number of reads per gene. In differential expression analysis, one is interested in reconstructing the level of expression of particular genes, which are the same in all cells, while in RepSeq one is interested in the number of cells expressing a given clonotype.
Thus, in RepSeq the number of transcripts will depend on the number of cells carrying that clonotypes, but also on their expression level, which is assumed to be clonotype-independent but noisy. There are thus three levels of noise in RepSeq: cell sampling noise, expression noise, and mRNA capture noise. By constrast in differential expression there is expression noise, cell-to-cell variability, and capture noise. These sources of noises combine in a different manner than in RepSeq.

edgeR \cite{Robinson2008}, a classical differential expression analysis software, proceeds by learning a noise model using a negative binomial model for expression noise from two identical conditions. Then, comparing RNA-seq data from two datasets, it evaluates a p-value corresponding to the probability that the observed difference in expression between the two datasets has occured just because of noise. We applied edgeR treating each clonotype as a separate gene.}

\subsection*{Obtaining diversity estimates from the clone frequency density}\label{sec:infer_div}
For a set of clone frequencies, $\{f_i\}_{i=1}^{N}$, the Hill family of diversities are obtained from the R\'enyi entropies, as $D_\beta=\exp H_\beta$, with $H_\beta=\frac{1}{1-\beta}\ln \left[ \sum_{i=1}^N f_i^{\beta}\right]$. We use $\rho(f)$ to compute their ensemble averages over $f$, again under the assumption that the joint distribution of frequencies factorizes. We obtain an estimate for $D_0=N$ using the model-derived expression, $N_{\textrm{obs}}+P(n=0)N=N$, where $N_{\textrm{obs}}$ is the number of clones observed in one sample, and $P(n=0)=\int_{f_{\textrm{min}}}^1 P(n=0|f)\rho(f)\textrm{d}f$. For $\beta=1$, we compute $\exp (N\langle -f\log f \rangle_{\rho(f)})$ and for $\beta=2$, we use $1/\left(N\langle f^2\rangle_{\rho(f)}\right)$.

\subsection*{Differential model sampling}\label{sec:diffexpr_sampling}
Since the differential expression model involves expansion and contraction in the test condition, some normalization in this condition is needed such that it produces roughly the same total number of cells as those in the reference condition, consistent with the observed data. One approach (the one taken below) is to normalize at the level of clone frequencies. 
Here, we instead perform the inefficient but more straightforward procedure of sampling all $N$ clones and discarding those clones for which $(n,n^{\prime})=(0,0)$. A slight difference in the two procedures is that $N_{\textrm{obs}}$ is fixed in the former, while is stochastic in the latter.

The frequencies of the first condition, $f_i$, are sampled from $\rho(f)$ until they sum to 1 (i.e. until before they surpass 1, with a final frequency added that takes the sum exactly to 1). An equal number of log-frequency fold-changes, $s_i$, are sampled from $\rho(s)$. The normalized frequencies of the second condition are then $f'_i=f_ie^{s_i}/\sum_j f_je^{s_j}$.  Counts from the two conditions are then sampled from $P(n|f)$ and $P(n^{\prime}|f')$, respectively. Unobserved clones, i.e. those with $(n,n^{\prime})=(0,0)$, are then discarded.

\subsection*{Inferring the differential expression prior}\label{sec:EM}
To learn the parameters of $\rho(s)$, we performed a grid search, refined by an iterative, gradient-based search to obtain the maximum likelihood. {\ch We tested different forms of prior shown in \cref{tab:prior}}.

\begin{table*}[htbp]
	\centering
	\normalsize
	\begin{tabular}{l|l|l}
		\hline
		      & Form of prior & Average data likelihood\\
		\hline
			full asymmetric exp.   & $(1-\alpha)\delta(s-s_0) + \alpha\Theta(s-s_0)e^{-\frac{s-s_0}{\bar{s}}}/\bar{s}$ & -1.894891\\
		\hline
			symmetric exp.        & $(1-\alpha)\delta(s-s_0) + \alpha e^{-\frac{|s-s_0|}{\bar{s}}}/2\bar{s}$ & -1.894303\\
		\hline
			centered Gaussian & $(1-\alpha)\delta(s-s_0) + \alpha e^{-\frac{(s-s_0)^2}{2\sigma^2}}/\sqrt{2\pi}\sigma$ & -1.894723 \\
		\hline
			off-centered Gaussian & $(1-\alpha)\delta(s-s_0) + \alpha e^{-\frac{(s-(s_0+s_1))^2}{2\sigma^2}}/\sqrt{2\pi}\sigma$ & -1.895101 ($s_1\geq 0.1$)\\
			\hline
	\end{tabular}
	\caption{Likelihoods for alternative forms of log-change prior distribution (donor S2; day 0-day-15). Note that the off-centered Gaussian was strictly off-centered, explaining its lower performance relative to the centered Gaussian despite having more degrees of freedom.}
	\label{tab:prior}
\end{table*}

For a more formal approach, expectation maximization (EM) can be employed when tractable. Here in a simple setting, we demonstrate this approach of obtaining the optimal parameter estimates from the data by calculating the expected log likelihood over the posterior and then maximizing with respect to the parameters. In practice, we first perform the latter analytically and then evaluate the former numerically. We choose a symmetric exponential as a tractable prior for this purpose:
\begin{align}
	\rho_{\rm exp}(s|\bar{s})=e^{-|s|/\bar{s}}/2\bar{s}
\end{align}
with $\bar{s}>0$, and no shift, $s_0=0$. The expected value of the log likelihood function, often called the Q-function in EM literature, is 
 \begin{align}
 Q(\bar{s}|\bar{s}')=\sum_{i=1}^{N_\textrm{\textrm{obs}}}\int_{-\infty}^{\infty}\mathrm{d}s\rho(s|n_i,n_i^{\prime},\bar{s}')\log \left[P(n_i,n_i^{\prime},s|\bar{s})\right]\;,
 \end{align}
 where $\bar{s}'$ is the current estimate.
 Maximizing $Q$  with respect to $\bar{s}$ is relatively simple since $\bar{s}$ appears only in $\rho_{\rm exp}(s|\bar{s})$  which is a factor in $P(n,n^{\prime},s|\bar{s})$. For each $s$,
 \begin{align}
 \frac{\partial \log \left[\rho_{\rm exp}(s|\bar{s}))\right] }{\partial\bar{s}} &=\frac{1}{\rho_{\rm exp}(s|\bar{s})} \frac{\partial\rho_{\rm exp}(s|\bar{s})}{\partial\bar{s}}\\&=\frac{|s|-\bar{s}}{\bar{s}^2}\;,
 \end{align}
so that $  \frac{\partial Q(\bar{s}|\bar{s}')}{\partial\bar{s}}=\sum_{i=1}^{N_\textrm{obs}}\int_{-\infty}^{\infty}\mathrm{d}s\rho(s|n_i,n_i^{\prime},\bar{s}')\frac{\partial \log \left[\rho_{\rm exp}(s|\bar{s}))\right] }{\partial\bar{s}} =0$ implies
\begin{align}
  \sum_{i=1}^{N_\textrm{obs}}\int_{-\infty}^{\infty}\mathrm{d}s\rho(s|n_i,n_i^{\prime},\bar{s}')\frac{|s|-\bar{s}^*}{\bar{s}^{*2}} =0
\end{align}
so that $\bar{s}^*=\frac{1}{N_\textrm{\textrm{obs}}}\sum_{i=1}^{N_\textrm{obs}}\bar{s}_{(n_i,n_i^{\prime})}$, where 
\begin{align}
\bar{s}_{(n,n^{\prime})}=\int_{-\infty}^{\infty}\mathrm{d}s|s|\rho(s|n,n^{\prime},\bar{s}').
\end{align}
The latter integral is computed numerically from the model using $\rho(s|n,n^{\prime},\bar{s}')=P(n,n^{\prime},s|\bar{s}')/\int_{-\infty}^{\infty}P(n,n^{\prime},s|\bar{s}')\mathrm{d}s	$. $Q$ is maximized at $\bar{s}=\bar{s}^*$ since  $ \frac{\partial^2 \log \left[\rho_{\rm exp}(s|\bar{s}))\right] }{\partial\bar{s}^2}\bigg|_{\bar{s}=\bar{s}^*}=-\bar{s}^{*-2} <0$. Thus, we update $\rho_{\rm exp}(s|\bar{s})$ with 
$\bar s\leftarrow\bar{s}^*$.
The number of updates typically required for convergence was small.

The constraint of equal repertoire size, $Z^\prime=Z$ can be satisfied with a suitable choice of the shift parameter, $s_0$, in the prior for differential expression, $\rho_{s}(s)$, namely $s_0=-\ln Z^\prime/Z$. The latter arises from the coordinate transformation $s\leftarrow\Delta s+s_0$, and adds a factor of $e^{s_0}$ to all terms of $Z^\prime$.

\section*{Acknowledgements}
MPT would like to thank M. Pogorelly for providing the R code used to obtain the EdgeR estimates in \citep{Pogorelyy12704}.  This work was supported by the European Research Council Consolidator Grant n. 724208.

\bibliographystyle{pnas}

\begin{thebibliography}{10}

\bibitem{Benichou2011}
Benichou J, Louzoun Y
\newblock (2011) {Rep-Seq : uncovering the immunological repertoire through
  next-generation sequencing}.
\newblock \emph{Immunology} pp 183--191.

\bibitem{Glanville2017}
Glanville J, {et~al.}
\newblock (2017) {Identifying specificity groups in the T cell receptor
  repertoire}.
\newblock \emph{Nature} 547:94--98.

\bibitem{Chu2019}
Chu ND, {et~al.}
\newblock (2019) {Longitudinal immunosequencing in healthy people reveals
  persistent T cell receptors rich in highly public receptors}.
\newblock \emph{BMC Immunology} 20:1--12.

\bibitem{Love2014}
Love MI, Huber W, Anders S
\newblock (2014) {Moderated estimation of fold change and dispersion for
  RNA-seq data with DESeq2.}
\newblock \emph{Genome biology} 15:550.

\bibitem{Robinson2008}
Robinson MD, Smyth GK
\newblock (2008) {Small-sample estimation of negative binomial dispersion ,
  with applications to SAGE data}.
\newblock \emph{Biostatistics} 9:321--332.

\bibitem{Pogorelyy12704}
Pogorelyy MV, {et~al.}
\newblock (2018) {Precise tracking of vaccine-responding T cell clones reveals
  convergent and personalized response in identical twins}.
\newblock \emph{Proceedings of the National Academy of Sciences}
  115:12704--12709.

\bibitem{Breda2019}
Breda J, Zavolan M, Nimwegen EV
\newblock (2019) {Bayesian inference of the gene expression states of single
  cells from scRNA-seq data}.
\newblock \emph{bioRxiv} p 2019.12.28.889956.

\bibitem{Mora2016e}
Mora T, Walczak A
\newblock (2018) in \emph{Systems Immunology}, eds{} Das JD, Jayaprakash C
\newblock (CRC Press), pp 185--199.

\bibitem{Gerritsen_thesis}
Gerritsen B
\newblock (2018) Ph.D. thesis (Utrecht University).

\bibitem{Greef2019}
Greef PCD, {et~al.}
\newblock (2019) {The naive T-cell receptor repertoire has an extremely broad
  distribution of clone sizes}.
\newblock \emph{bioRxiv:691501}.

\bibitem{Zarnitsyna2013}
Zarnitsyna VI, Evavold BD, Schoettle LN, Blattman JN, Antia R
\newblock (2013) {Estimating the diversity, completeness, and cross-reactivity
  of the T cell repertoire}.
\newblock \emph{Frontiers in Immunology} 4:485.

\bibitem{Heather2017}
Heather JM, Ismail M, Oakes T, Chain B
\newblock (2017) {High-throughput sequencing of the T-cell receptor repertoire:
  pitfalls and opportunities}.
\newblock \emph{Briefings in Bioinformatics} 19:554--565.

\bibitem{Robinson2010}
Robinson MD, Mccarthy DJ, Smyth GK
\newblock (2010) {edgeR : a Bioconductor package for differential expression
  analysis of digital gene expression data}.
\newblock \emph{Bioinformatics} 26:139--140.

\bibitem{Qi2014}
Qi Q, {et~al.}
\newblock (2014) {Diversity and clonal selection in the human T-cell
  repertoire.}
\newblock \emph{Proceedings of the National Academy of Sciences of the United
  States of America} 111:13139--44.

\bibitem{Lythe2016}
Lythe G, Callard RE, Hoare RL, Molina-Par{\'{i}}s C
\newblock (2016) {How many TCR clonotypes does a body maintain?}
\newblock \emph{Journal of Theoretical Biology} 389:214--224.

\bibitem{Jenkins2010}
Jenkins MK, Chu HH, McLachlan JB, Moon JJ
\newblock (2009) {On the composition of the preimmune repertoire of T cells
  specific for Peptide-major histocompatibility complex ligands.}
\newblock \emph{Annual review of immunology} 28:275--294.

\bibitem{Mora2019}
Mora T, Walczak AM
\newblock (2019) {How many different clonotypes do immune repertoires contain?}
\newblock \emph{Current Opinion in Systems Biology} 18:104--110.

\bibitem{Boer1993}
de~Boer RJ, Perelson ASA
\newblock (1993) {How diverse should the immune system be}.
\newblock \emph{Proceedings of the Royal Society B: Biological Sciences}
  252:171.

\bibitem{Mayer2019}
Mayer A, Zhang Y, Perelson AS, Wingreen NS
\newblock (2019) {Regulation of T cell expansion by antigen presentation
  dynamics}.
\newblock \emph{Proceedings of the National Academy of Sciences of the United
  States of America} 116:5914--5919.

\bibitem{Anders2010}
Anders S, Huber W
\newblock (2010) {Differential expression analysis for sequence count data.}
\newblock \emph{Genome biology} 11:R106.

\bibitem{Ralph2016a}
Ralph DK, Matsen FA
\newblock (2016) {Likelihood-Based Inference of B Cell Clonal Families}.
\newblock \emph{PLoS Computational Biology} 12:1--28.

\bibitem{Minervina2019}
Minervina A, Pogorelyy M, Mamedov I
\newblock (2019) {TCR and BCR repertoire profiling in adaptive immunity}.
\newblock \emph{Transplant International} pp 0--2.

\bibitem{Kadoki2017}
Kadoki M, {et~al.}
\newblock (2017) {Organism-Level Analysis of Vaccination Reveals Networks of
  Protection across Tissues}.
\newblock \emph{Cell} 171:398--413.e21.

\bibitem{Nourmohammad2019}
Nourmohammad A, Otwinowski J, {\L}uksza M, Mora T, Walczak AM
\newblock (2019) {Fierce Selection and Interference in B-Cell Repertoire
  Response to Chronic HIV-1}.
\newblock \emph{Molecular Biology and Evolution} 36:2184--2194.

\end{thebibliography}

\vskip 5cm

\beginsupplement


\begin{figure}[h!]
\includegraphics{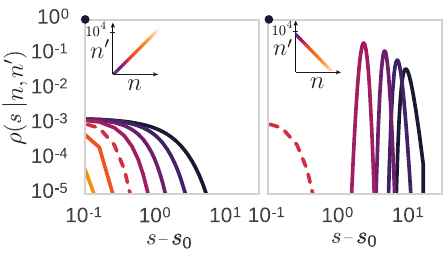}
\caption{
Posteriors of the learned model, $\rho(s|n,n^{\prime})$ over pairs $(n,n^{\prime})$ for $n^{\prime}=n$, with $n$ varying over a logarithmically-spaced set of counts (left), and for $n^{\prime}$ given by the reverse order of this set (right). The black dot in both plots denotes the contribution of the non-responding component, $\propto \delta(s-s_0)$, to the posterior.
(Parameters: $N=10^6$, $\epsilon=10^{-2}$.)
\label{fig:posterior_var}}
\end{figure}

\end{document}